\documentclass[10pt,conference]{IEEEtran}
\IEEEoverridecommandlockouts

\usepackage{cite}
\usepackage{amsmath,amssymb,amsfonts}
\usepackage{algorithmic}
\usepackage{graphicx}
\usepackage{textcomp}
\usepackage{xcolor}
\def\BibTeX{{\rm B\kern-.05em{\sc i\kern-.025em b}\kern-.08em
    T\kern-.1667em\lower.7ex\hbox{E}\kern-.125emX}}

\usepackage{booktabs}
\usepackage{multirow}
\usepackage{tabularray}
\usepackage{rotating}
\usepackage{adjustbox}
\usepackage[misc]{ifsym}
\usepackage{subcaption}
\usepackage{tikz}

\usepackage[colorlinks=true]{hyperref}

\definecolor{lightgray}{gray}{0.96} 

\begin{document}


\title{Evolutionary Multi-Objective Fusion of\\
Deepfake Speech Detectors
\thanks{This work was supported by the BUT internal project FIT-S-26-901 and the Czech Science Foundation project 24-10990S. Computational resources were provided by the e-INFRA CZ project (ID:90254), supported by the Ministry of Education, Youth and Sports of the Czech Republic. The authors acknowledge the use of generative tools for assistance with grammar and text refinements.}
}

\author{\IEEEauthorblockN{Vojtěch Staněk\textsuperscript{\Letter}, Martin Perešíni, Lukáš Sekanina, Anton Firc, Kamil Malinka}
\IEEEauthorblockA{\textit{Faculty of Information Technology}, \textit{Brno University of Technology} \\
Božetěchova 1/2, 612 00 Brno, Czech Republic \\
\{istanek, iperesini, sekanina, ifirc, malinka\}@fit.vut.cz}}



\maketitle

\begin{abstract}
    While deepfake speech detectors built on large self-supervised learning (SSL) models achieve high accuracy, employing standard ensemble fusion to further enhance robustness often results in oversized systems with diminishing returns. 
    To address this, we propose an evolutionary multi-objective score fusion framework that jointly minimizes detection error and system complexity.
    We explore two encodings optimized by NSGA-II: binary-coded detector selection for score averaging and a real-valued scheme that optimizes detector weights for a weighted sum.
    Experiments on the ASVspoof 5 dataset with 36 SSL-based detectors 
    show that the obtained Pareto fronts outperform simple averaging and logistic regression baselines.
    The real-valued variant achieves 2.37\% EER (0.0684 minDCF) and identifies configurations that match state-of-the-art performance while significantly reducing system complexity, requiring only half the parameters.
    Our method also provides a diverse set of trade-off solutions, enabling deployment choices that balance accuracy and computational cost.
\end{abstract}

\begin{IEEEkeywords}
Evolutionary Multi-objective Optimization, Deepfake Speech Detection, Ensemble Learning, NSGA-II
\end{IEEEkeywords}

\section{Introduction}

The rapid evolution of high-fidelity speech synthesis has enabled the creation of highly realistic deepfakes, which can impersonate individuals with remarkable fidelity, thereby posing a threat to the security of voice biometric systems~\cite{Firc2, SalkoFaceDeepfakes}. In response, the research community has developed increasingly sophisticated deepfake speech detectors~\cite{AASIST}, with recent advancements utilizing large-scale self-supervised learning (SSL) models to achieve state-of-the-art performance~\cite{wang24_asvspoof}.

However, this pursuit of detection accuracy has come at a high cost: modern detectors have grown exponentially in size and computational complexity~\cite{chen24_asvspoof}. To further enhance robustness and generalize across diverse attacks, researchers often resort to ensemble learning, i.e., fusing multiple detectors~\cite{combei24_asvspoof}. However, standard fusion techniques typically optimize a single performance metric (e.g., Equal Error Rate or Decision Cost~\cite{wang24_asvspoof}). This single-objective focus inevitably leads to systems that accumulate models, including redundant or marginally contributing ones. Additionally, adding more and more models yields diminishing returns~\cite{xu24_asvspoof}.

Consequently, a critical methodology gap exists: systematically balancing detection performance against system efficiency, leaving no clear path for constructing compact but high-performing systems. To address this, we propose a multi-objective evolutionary algorithm (MOEA) capable of selecting the most suitable subset of detectors from a pool of detectors, thus balancing detection quality and detector complexity (model size). On standard benchmark problems, the evolved ensemble of detectors shows superior performance when compared to state-of-the-art approaches.


The main \textbf{contributions} can be summarized as follows:

\begin{itemize} 
    \item We introduce a novel application of MOEA in the domain of deepfake speech detection. Unlike traditional single-objective methods that seek a single optimal point, our approach simultaneously optimizes conflicting objectives (maximizing detection performance while minimizing system complexity) to produce a comprehensive Pareto front of suitable solutions.

    \item We propose a MOEA optimization framework utilizing the NSGA-II algorithm, and compare two problem encodings for MOEA (a binary-coded selection of detectors and a weighting-based selection of detectors in ensembles).

    \item We demonstrate that this systematic approach yields fusions that outperform current state-of-the-art benchmarks by identifying configurations that achieve either lower Equal Error Rate (EER) or comparable performance with significantly higher resource efficiency.
\end{itemize}





\section{Background}


\subsection{Deepfake Speech Detection}








The increasing malicious use of deepfakes motivates the development of deepfake speech detectors~\cite{Firc2}. Commonly used approaches generally employ three stages: \textit{feature extraction}, \textit{model training}, and finally \textit{classification}~\cite{Almutairi2022}. Analysis of the 2024 ASVspoof 5 challenge~\cite{wang24_asvspoof} 
revealed that current systems employ architectures based on deep neural networks~\cite{wang24_asvspoof}. Current architectures utilize pre-trained Self-Supervised Learning (SSL) models such as Wav2Vec2~\cite{Wav2Vec2} and WavLM~\cite{WavLM}. 
The rich features extracted by SSL models are further processed and pooled~\cite{scdf}, most commonly by Graph Attention Networks (GAT)~\cite{tak21_interspeech} from the AASIST framework~\cite{AASIST}. Alternatively, modern architectures utilize Multi-Head Factorized Attention pooling (MHFA)~\cite{peng2022_mhfa} or a Sensitive Layer Selection (SLS) classifier~\cite{SLS}.

\begin{figure*}[t]
    \centering
    \includegraphics[width=\linewidth]{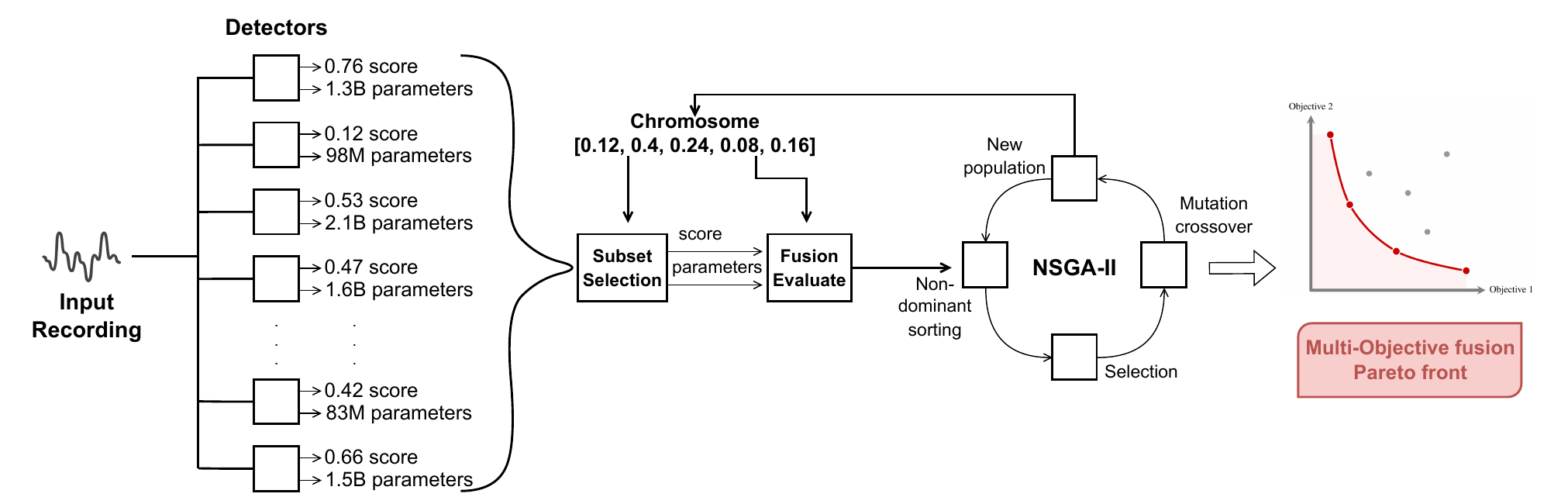}
    \caption{Overview of the proposed evolutionary multi-objective fusion framework. The system processes input recordings through a pool of base detectors to obtain scores and parameter counts. A candidate fusion is constructed and evaluated based on a randomly initialized chromosome. NSGA-II then iteratively optimizes the fusion configuration (i.e., the chromosome) to simultaneously minimize the Equal Error Rate (EER) and system complexity (i.e., the number of parameters), ultimately producing a Pareto front of optimal trade-off solutions.}
    \label{fig:framework}
    \vspace{-0.5em}
\end{figure*}

\subsection{Ensemble Learning and Classifier Fusion}
\label{subsec:background_fusion}

Model fusion, also known as ensemble learning, is commonly used in many machine learning areas~\cite{wang24_asvspoof} 
to improve robustness and reduce variance across operating conditions~\cite{Chettri2019}. 
In practice, fusion is most often carried out at the \textit{score level}, which treats each detector as a black box that outputs a detection score~\cite{stourbe24_asvspoof}. This makes it straightforward to combine heterogeneous systems that use different extractors, classifiers, or even training objectives, while avoiding costly retraining of models. It also allows for the reuse of existing systems and improving them by simply post-processing their scores~\cite{combei24_asvspoof}.

There are two main approaches for score-level fusion: \textit{aggregation} and \textit{learned combination}. Aggregation-based techniques (e.g., averaging~\cite{MHFA} or majority voting~\cite{majority_voting_fusion}) combine detector outputs through fixed mathematical operations, without requiring training, and offer a simple method for fusing multiple detector scores. On the other hand, learned combination methods (e.g., logistic regression~\cite{combei24_asvspoof}) learn optimal weights or mappings directly from detector scores to maximize detection performance, albeit at the cost of greater complexity and higher data requirements. 












\subsection{Evolutionary Machine Learning}

Under the umbrella of \emph{evolutionary machine learning}, the interaction between machine learning (ML) and evolutionary algorithms (EA) is typically categorized into three areas~\cite{EMLbook:2023}: (i) the use of EA to enhance ML methods; (ii) the use of ML techniques to enhance EA; and (iii) the application of EA to problems traditionally solved by standard ML approaches. In this paper, we focus on area (i), where EA can enhance multiple stages of the ML pipeline, including the construction and optimization of ensembles of ML models.

Evolutionary approaches to ensemble design have been extensively studied in the literature, particularly for neural networks as base ML models~\cite{LYH:2000}, and across a wide range of applications. EA-driven enhancement of ensembles has been investigated at different levels of problem complexity. Examples include: using EA to identify an optimal subset of candidate ML models~\cite{Bosowski:2021}, optimizing pairing of ML models and input features~\cite{Kim:2008}, and applying genetic programming to evolve new base learners, including their input features and structure~\cite{Ying:2021}. Since ensemble construction and pruning are inherently multi-objective problems, multi-objective evolutionary algorithms (MOEAs) have been widely applied to solve them. The final set of trade-off solutions is represented as a Pareto front of non-dominated ensembles. To approximate this front, domination-based methods (e.g., NSGA-II~\cite{nsga2}, NSGA-III~\cite{nsga3}) and decomposition-based approaches (e.g., MOEA/D~\cite{Zhang:moea:d:2007}) are commonly employed.

\section{Proposed Multi-Objective Fusion Approach}


Most deepfake detection methods use a single-objective selection of detectors.
The design of an efficient score-level fusion system for deepfake speech detection can be formally framed as a multi-objective optimization problem. Our primary goal is to find a fusion configuration that effectively balances two objectives: maximizing detection performance and minimizing system complexity. Additionally, we aim to generate multiple solutions that cover different trade-offs between performance and complexity, allowing for the selection of the most suitable configuration for a given scenario. The overview of the proposed framework is presented in~\autoref{fig:framework}. 

\begin{figure}[t]
    \centering
    \includegraphics[width=\linewidth]{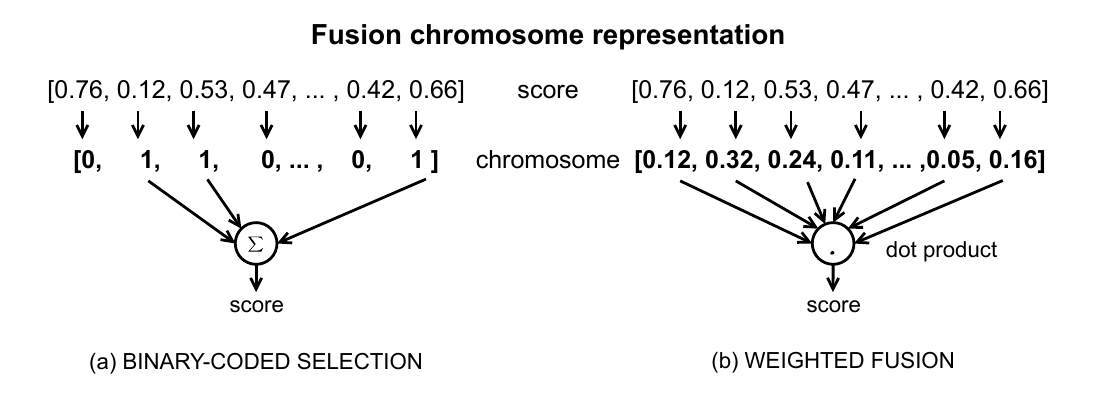}
    \caption{Chromosome representations for the two variants.}
    \label{fig:chromosome}
\end{figure}

For multi-objective optimization, evolutionary algorithms such as the established NSGA-II~\cite{nsga2} can be employed. Therefore, we investigate two variants of our proposed multi-objective fusion based on the problem coding and solution representation, as shown in \autoref{fig:chromosome}.

\begin{enumerate}
    \item \textbf{Binary-coded detector selection}
    

    In this variant, the chromosome $\mathbf{c} \in \{0, 1\}^D$ is a binary vector of length $D$ (where $D$ is the number of base detectors). Each gene $c_i \in \mathbf{c}$ corresponds to the $i$-th base detector. A value of $c_i = 1$ indicates the inclusion of the detector, whereas $c_i = 0$ indicates that the detector is not part of the fusion. The final fusion score $s_{final}$ is computed as the average of the scores from the selected detectors:
    
    \begin{equation*}
        s_{final} = \frac{\sum_{i=1}^{D} c_i \cdot s_i}{\sum_{i=1}^{D} c_i}
    \end{equation*}
    
    where $s_i$ is the output score of the $i$-th detector.
    
    \item \textbf{Learning detector weights (real-valued)}
    

    This variant focuses on learning the fusion weights for each fused detector. The chromosome is a real-valued vector $\mathbf{c} \in [0, 1]^D$, where each gene $c_i$ represents a non-negative weight of the $i$-th detector. To effectively explore the objective of low system complexity, we introduce a cut-off threshold $W$. Detectors with weights falling below this threshold are considered to have a negligible impact on the fusion; their weights are set to zero, and they are excluded from the final fusion. This mechanism enables the algorithm to prune low-contributing detectors, allowing for a targeted search for solutions that achieve a reasonable trade-off between the objectives. The remaining weights are then normalized to sum to 1:
    \begin{equation*}
        w_i = \frac{\hat{c}_i}{\sum_{j=1}^{D} \hat{c}_j}, \quad \text{where } \hat{c}_i =
        \begin{cases}
        c_i & \text{if } c_i \ge W \\
        0 & \text{if } c_i < W
        \end{cases}
    \end{equation*}
    
    The final fusion score is computed as the weighted sum of the individual detector scores $s_i$:
    
    \begin{equation*}
        s_{final} = \sum_{i=1}^{N} w_i \cdot s_i
    \end{equation*}
\end{enumerate}

\subsection{Fusion objectives}

The first objective $f_1$ is to maximize detection performance, i.e., minimize error rates and false classifications. In biometric and spoof detection tasks, two primary types of classification errors are commonly considered:

\begin{itemize}
    \item False Acceptance Rate\footnote{Also known as APCER -- Attack Presentation Classification Error Rate.}\textbf{} (FAR): the proportion of deepfake samples incorrectly classified as bonafide.
    \item False Rejection Rate\footnote{Also known as BPCER -- Bonafide Presentation Classification Error Rate.} (FRR): the proportion of bonafide samples incorrectly classified as deepfake.
\end{itemize}

The Equal Error Rate (EER) is defined as the point where these two error rates are equal: 
$$EER = FAR(\tau^*) = FRR(\tau^*)$$ 
where $\tau^*$ is the threshold, where $FAR = FRR$. EER is widely adopted in biometric and spoof detection tasks as it is threshold-independent and reflects the trade-off between FAR and FRR. A lower EER indicates better separation between bonafide and deepfake trials, making it a robust and interpretable metric for detection performance.

The second objective $f_2$ is to minimize system complexity. For this, we use the number of parameters of the resulting system as a proxy for system size, memory demands, and computational requirements. Intuitively, bigger and more complex systems should be able to learn more complex relationships or representations and achieve better performance (lower EER) at the cost of computational intensity (more parameters, i.e., more operations to compute).

Therefore, in our case of efficiently fusing multiple deepfake speech detectors, we aim to minimize the following two objective functions:

$$f_1(\mathbf{c}) = EER(\mathbf{c})$$
$$f_2(\mathbf{c}) = Number\_of\_Parameters(\mathbf{c})$$

Here, \textbf{c} represents a candidate fusion configuration, which can be either a binary-coded selection of base detectors or a real-valued vector of detector weights, depending on the variant described above. The $f_1$ objective minimizes the EER of the fused system (i.e., maximizes detection performance), while $f_2$ minimizes the number of parameters of the fused system as a proxy for system complexity.

\subsection{Evolutionary optimization mechanism}

To solve the presented multi-objective fusion problem, we employ the NSGA-II evolutionary algorithm~\cite{nsga2} for multi-objective optimization to discover the set of Pareto-optimal solutions, i.e., a well-covered front of solutions with different trade-offs between performance and system complexity. 
In our implementation, the genetic operators are adapted to the encoding of each fusion variant. For both variants, we employed a uniform crossover to create offsprings. The binary variant then uses bit-flip mutation to explore new solutions. The real variant uses a standard polynomial mutation~\cite{polynomial_mutation}, which introduces an additional hyperparameter $\eta_m$ that shapes the mutation distribution\footnote{A larger $\eta_m$ value creates a more narrow distribution, i.e., smaller mutations, while a lower $\eta_m$ value yields wider, more exploratory mutations.}, in addition to the cut-off threshold $W$. Weights smaller than $W$ are set to 0 and excluded from the final fusion. Finally, the mutated weight vector is normalized so that its elements sum to 1.

To select individuals for the next generation, we use a standard binary tournament selection approach, where the individual from the lower-ranked (superior) non-dominated front is selected. If both individuals come from the same front, the one with a higher crowding distance is selected, following the same procedure as in the original NSGA-II paper~\cite{nsga2}.

To quantify the quality and diversity of the generated Pareto front, we employ the Hypervolume (HV) metric. HV measures the volume of the objective space dominated by the best non-dominated set relative to a defined reference point. A higher HV indicates superior performance in both convergence (closer to the optimal trade-off) and diversity (better spread of solutions)~\cite{nsga2}.
Importantly, the proposed evolutionary approach can be easily extended beyond the two defined objectives. Additional objectives can stem from deployment constraints; for example, inference time or memory requirements may be crucial in certain scenarios. Similarly, a timely topic in deepfake speech detection is the robustness against adversarial attacks~\cite{rahibi24_adversarial}. 

\section{Experiment setup}

Our fusion pool consists of 36 distinct deepfake detectors. The models are constructed from four pre-trained SSL models (HuBERT~\cite{HuBERT}, Wav2Vec2~\cite{Wav2Vec2}, XLS-R~\cite{xlsr}, WavLM~\cite{WavLM}), each available in three sizes. The rich features extracted by these SSL models are then processed by three pooling and classifier architectures (AASIST~\cite{AASIST}, MHFA~\cite{MHFA}, SLS~\cite{SLS}). 

For our experiments, we utilized the ASVspoof 5 da\-ta\-set~\cite{wang24_asvspoof}, a well-established benchmark in the deepfake speech detection research area. It provides standardized evaluation protocols and data for comparing deepfake speech detectors. As usual, we used disjoint parts of the dataset during the experiments: the train split for training the base detectors, the dev split for training the fusions, and the eval split for the final evaluation of the trained detectors and fusions. As previously mentioned, we used EER as the performance metric, and the number of parameters of the final fusion serves as a proxy for measuring system complexity.

\subsection{Base detectors}

We trained the models for ten epochs on the training split of the ASVspoof 5 dataset, while keeping the pre-trained SSL model frozen; i.e., we optimized only the classifier weights. We used the default Adam optimizer from PyTorch, along with a simple Cross-entropy loss function, with a learning rate of $10^{-3}$ and a batch size ranging from 4 (for large models) to 32 (for smaller models).

The parameter count for these detectors ranges from 95M to 2.24B, while their individual EERs span from 4.79\% to 19.32\%, covering a wide range of sizes and detection performances. A detailed summary of these individual performances and their respective parameter counts is provided in \autoref{tab:individual_eers}. We did not aim to fine-tune or extensively optimize the performance of these underlying base detectors to achieve state-of-the-art results as standalone systems. Instead, our primary goal was to create a diverse pool of detectors with a broad range of performance and sizes to evaluate our proposed fusion solution. 

\subsection{Baseline fusions}

We created several baseline fusions to put the proposed method in perspective. We selected three suitable manual fusions, which average the scores of the selected detectors:

\begin{enumerate}
    \item The \textit{Light} fusion combines the best variants of the three classifier architectures (AASIST, MHFA, and SLS). Averaging the scores of XLS-R 1B (AASIST), XLS-R 2B (SLS), and WavLM large (MHFA), marked by bold text in \autoref{tab:individual_eers}, results in a fusion of 3.52B parameters and EER of 3.95\%.
    \item The \textit{Heavy} fusion combines the best classifier architecture of each SSL model, marked by italicized text in \autoref{tab:individual_eers}, resulting in an EER of 3.35\% with 6.21B parameters.
    \item Averaging the scores of all 36 base detectors does not yield better results, as the worse-performing models contribute equally as the better-performing ones, yielding a large fusion of 18.56B parameters with 3.44\% EER.
\end{enumerate}

Secondly, we implemented the Logistic Regression classifier on top of the base detector scores. The performance of this full fusion of all 36 models yields an EER of 2.57\% with a total of 18.56B parameters. Furthermore, we progressively removed the worst-performing or least-contributing individual base detectors (i.e., those with the lowest corresponding weights) from the fusion. In both cases, we observed a "long tail" effect, where the EER doesn't notably change until only 10 base detectors are present in the fusion, resulting in an EER of 2.63\% with a significant reduction in complexity to 5.5B parameters. Beyond this point, the EERs begin to rise more notably as additional models are removed. 

\begin{table}[tbp]
    \centering
    \caption{Equal Error Rate [\%] and number of parameters (in brackets) of individual deepfake speech detectors on the evaluation set of ASVspoof 5. The bold values indicate the systems used for the \textbf{Light} manual fusion, and italicized values indicate the systems used for the \textit{Heavy} manual fusion.}
    \label{tab:individual_eers}
    \SetTblrInner{rowsep=4pt}
    \begin{adjustbox}{max width=\linewidth}
        \begin{tblr}{
            width=\linewidth,
            colspec={l l r r r r r r},
            cell{1}{3,5,7} = {c=2}{c},
            cell{2,5,8,11}{1} = {r=3}{c},
            row{even} = {bg=lightgray},
            column{1} = {bg=white},
            vline{2,3,5,7} = {-}{},
            hline{3,4,6,7,9,10,12,13} = {2-8}{dotted},
            hline{2,5,8,11} = {-}{},
          }
        & \textbf{Model} & \textbf{AASIST} & & \textbf{MHFA} & & \textbf{SLS} & \\
        \begin{sideways}\textbf{HuBERT}\end{sideways} & base & 13.51\% & (95M) & \textit{10.82\%} & \textit{(96M)} & 12.52\% & (108M) \\
        & large & 8.06\% & (317M) & 7.74\% & (321M) & \textit{6.39\%} & \textit{(339M)} \\
        & x-large & \textit{8.11\%} & \textit{(964M)} & 8.45\% & (977M) & 8.88\% & (1.00B) \\
        \begin{sideways}\textbf{Wav2Vec2}\end{sideways} & base & 16.65\% & (95M) & \textit{12.42\%} & \textit{(96M)} & 15.86\% & (108M) \\
        & large & 19.32\% & (317M) & \textit{7.04\%} & \textit{(321M)} & 7.78\% & (339M) \\
        & LV60k & 8.64\% & (317M) & \textit{8.22\%} & \textit{(321M)} & 9.09\% & (339M) \\
        \begin{sideways}\textbf{XLS-R}\end{sideways} & 300M & 7.53\% & (317M) & \textit{5.34\%} & \textit{(321M)} & 6.68\% & (339M) \\
        & 1B & \textbf{5.56\%} & \textbf{(964M)} & 6.13\% & (977M) & \textit{5.09\%} & \textit{(1.00B)} \\
        & 2B & 7.41\% & (2.16B) & 6.28\% & (2.19B) & \textit{\textbf{4.79\%}} & \textit{\textbf{(2.24B)}} \\
        \begin{sideways}\textbf{WavLM}\end{sideways} & base & 12.30\% & (95M) & \textit{8.35\%} & \textit{(96M)} & 10.87\% & (108M) \\
        & base+ & 10.37\% & (95M) & \textit{8.65\%} & \textit{(96M)} & 9.26\% & (108M) \\
        & large & 7.03\% & (317M) & \textit{\textbf{5.12\%}} & \textit{\textbf{(321M)}} & 5.77\% & (340M) \\
        \end{tblr}
    \end{adjustbox}
\end{table}

\subsection{Fusions Optimized with NSGA-II}

We implemented the NSGA-II algorithm for the purposes of this experiment\footnote{Code: \url{https://github.com/Security-FIT/evolutionary_detector_fusion}}. To determine the optimal configuration for both encoding variants, we conducted a rigorous parameter sensitivity analysis. For hyperparameter tuning, we fixed a computational budget of 25,000 fitness evaluations (both objectives). We conducted 10 independent runs for each tested parameter combination, measuring the final mean HV to assess performance and ensure statistical robustness. We used a server equipped with an AMD EPYC 9124 16-Core (64 threads) CPU for all experiments.

For computing HV, we set the reference point to the worst observed performance from the base detectors: 20\% EER and 18.56B parameters (all detectors). This reference point defines relevant objective space, allowing HV to be normalized and compared across different generations and experiments.

\begin{figure}[htbp]
    \centering
    \includegraphics[width=\linewidth]{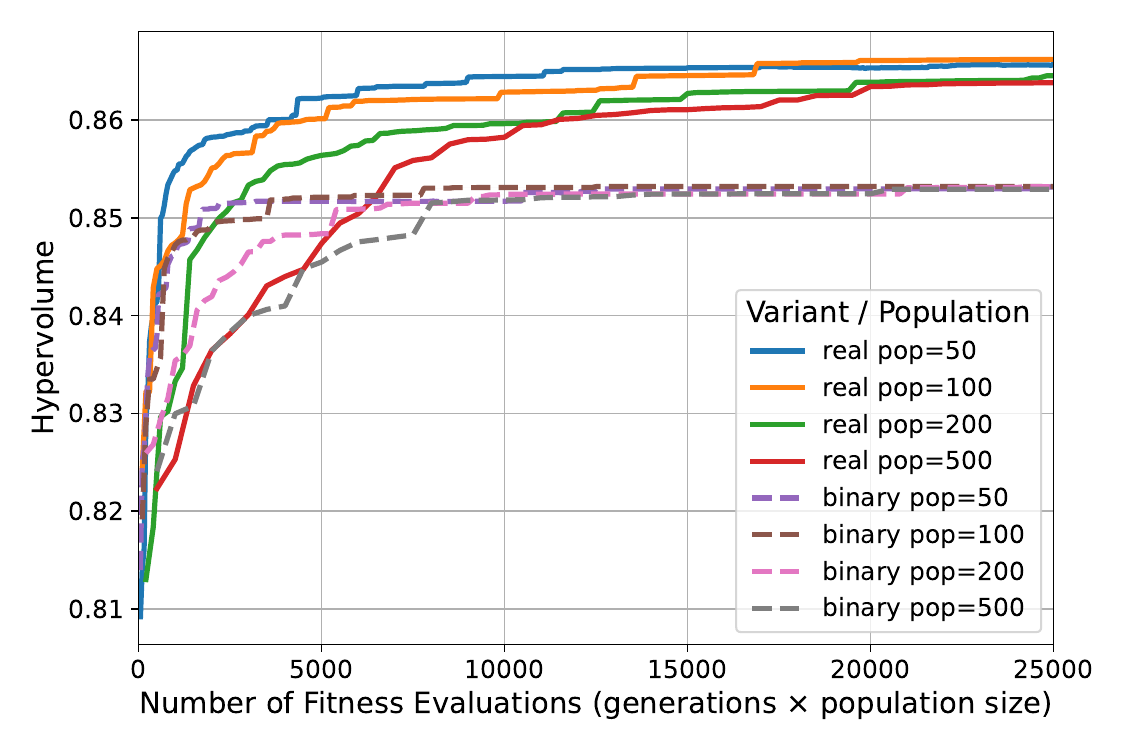}
    \caption{Impact of population size on HV convergence under fixed computational budget (best viewed in color). Averages of 10 runs are reported.}
    \label{fig:budget}
\end{figure}

First, we analyzed the trade-off between population size ($N$) and generation count ($G$). We fixed the other parameters at a mutation rate of 1/36, a crossover rate of 0.9, and $\eta_m = 5$. We evaluated population sizes of $N \in \{50, 100, 200, 500\}$; the results indicated that the population of 100 individuals provided the most stable HV growth and sufficient population diversity, as visible in \autoref{fig:budget}.

\begin{figure*}[htbp]
    \centering
    \includegraphics[width=\textwidth]{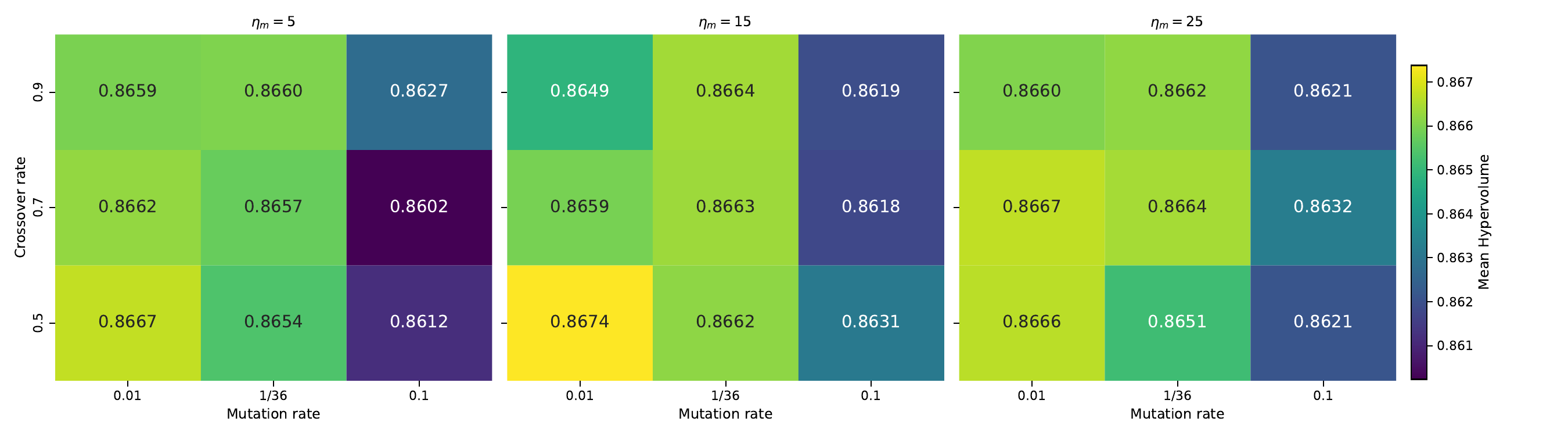}
    \caption{Parameter sensitivity analysis for the real-valued variant: mean HV vs. $p_m$ and $p_c$ for three distinct distribution indices ($\eta_m$) under a fixed computational budget of 25,000 fitness evaluations.}
    \label{fig:real_parameters}
    \vspace{-1em}
\end{figure*}

\begin{figure}[htbp]
    \centering
    \includegraphics[width=0.4\textwidth]{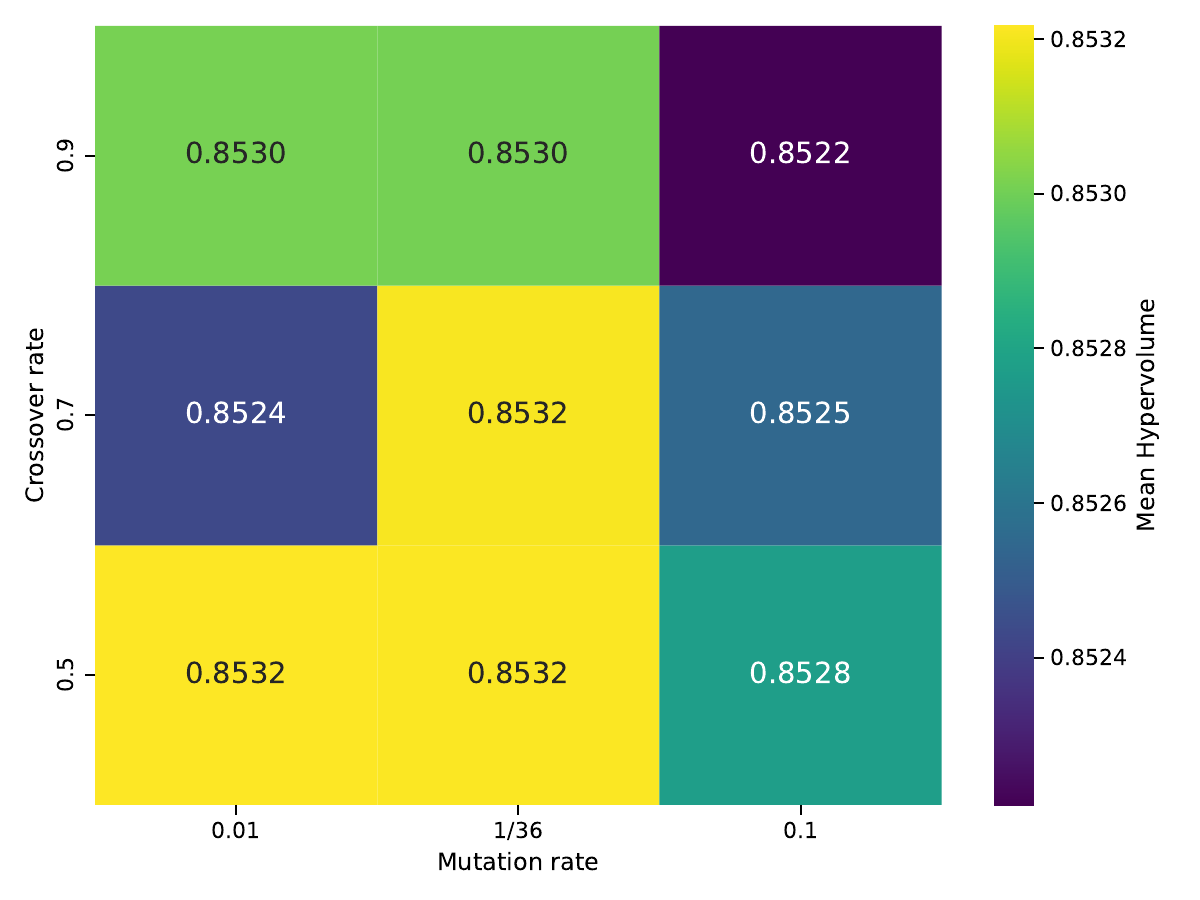}
    \caption{Parameter sensitivity analysis for the binary variant: mean HV vs. $p_m$ and $p_c$ under fixed computational budget.}
    \label{fig:binary_parameters}
    \vspace{-1.5em}
\end{figure}

Second, we performed a grid search for parameter tuning on crossover rate ($p_c \in \{0.5, 0.7, 0.9\}$), mutation rate ($p_m \in \{0.01, 1/36, 0.1\}$), and also the distribution index ($\eta_m \in \{5, 15, 25\}$) for the real-valued variant. Results are presented in \autoref{fig:real_parameters} and \autoref{fig:binary_parameters} on the next page.

\textbf{Binary-coded variant:} The highest mean HV was observed at lower mutation and crossover rates. The balanced configuration, with $p_c = 0.7$ and $p_m = 1/36$, converged the fastest, so we selected these parameters for further experiments.

\textbf{Real-valued variant:} A low mutation rate ($p_m = 0.01$) and crossover rate ($p_c=0.5$), combined with moderate $\eta_m = 15$, was observed to achieve the highest mean HV, so we selected these parameters for the real-valued variant.

A key aspect of the real-valued variant of learning the detector weights was the use of a cut-off threshold $W$ set to 0.001. This parameter determines which models are considered part of the fusion, as any weight value below this threshold is treated as zero. This specific value was chosen as a compromise, as a higher $W$ led to a search for fusions with a smaller number of models, while a lower $W$ resulted in larger fusions. The selected value of 0.001 enabled an effective search across the middle ground of the Pareto front. Therefore, depending on the desired use case, lowering $W$ should lead to a more intensive search in the area of large and higher performing fusions, while setting $W$ to a higher value should lead to exploration of smaller fusions.

\begin{figure}[htbp]
    \centering
    \includegraphics[width=\linewidth]{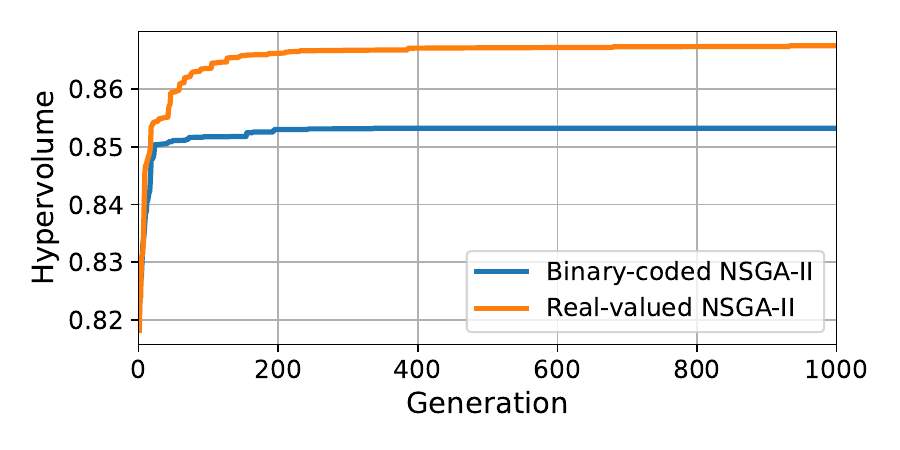}
    \caption{HV convergence -- averages of 10 runs are reported.}
    \label{fig:convergence}
\end{figure}

The final parameter setting is detailed in \autoref{tab:nsga_params}. Validation runs confirmed that the binary and real-valued variants converge within approximately 200 and 400 generations, respectively, as visible in \autoref{fig:convergence}. This difference in convergence speed can be attributed to the larger and more complex search space explored by the real-valued variant. 

\begin{table}[htbp]
\centering
\caption{Hyperparameters of the NSGA-II Fusions}
\begin{tblr}{
    colspec = {l|[dotted]c|[dotted]c},
    row{even} = {bg=lightgray},
    row{odd} = {bg=white},
    row{1} = {bg=white, font=\bfseries},
    hline{1,11} = {0.5pt, solid},
    hline{2} = {0.25pt, solid},
    rowsep = 3pt,
}
    Parameter                  & Binary Variant            & Real-Valued Variant \\
    Encoding                   & $\mathbf{c} \in \{0,1\}^D$ & $\mathbf{c} \in [0,1]^D$ \\
    Population Size ($N$)      & 100                       & 100 \\
    Selection                  & Binary Tournament         & Binary Tournament \\
    Crossover Type             & Uniform                   & Uniform \\
    Crossover Rate ($p_c$)     & 0.7                       & 0.5 \\
    Mutation Type              & Bit-flip                  & Polynomial \\
    Mutation Rate ($p_m$)      & $1/36$                    & $0.01$ \\
    Distribution Index ($\eta_m$) & --                     & 15 \\
    Cut-off Threshold ($W$)    & --                        & 0.001
\end{tblr}
\label{tab:nsga_params}
\vspace{-1em}
\end{table}

To ensure the search begins with high-quality candidates and accelerates convergence, we used a hybrid population initialization strategy. The initial population was seeded with 37 individuals: firstly, the average fusion of all base detectors, and secondly, individuals corresponding to each of the 36 base detectors. The rest of the population was initialized at random. This heuristic ensures that the evolutionary process begins with a baseline performance at least as good as that of the best single detector, which has proven to be important for faster convergence and solution stability.

\textbf{Code Optimizations and Parallelism.} We replaced the iterative implementations of non-dominated sorting and crowding distance calculation with vectorized NumPy operations. For the real-valued variant with population $N = 1000$, this optimization reduced the computational cost of generation updates by 16$\times$ (on average from $5.821$s to $0.356$s). Furthermore, the most computationally intensive step, i.e., the objective fitness evaluation, was parallelized across all available cores using multi-threading to fully utilize the 64-thread architecture. The implementation is available in the GitHub repository mentioned above. For the final evaluation, we employed early stopping based on convergence to conserve computational resources: if the HV improvement remains below a threshold $\varepsilon = 10^{-5}$ for 30 generations, the algorithm is stopped. Together, these optimizations resulted in an average runtime of 303 seconds ($225$ generations) for a single NSGA-II run of the binary variant and 653 seconds ($433$ generations) for a single run of the real-valued variant.

\section{Results}

\begin{figure*}
    \centering
    \includegraphics[width=0.69\linewidth]{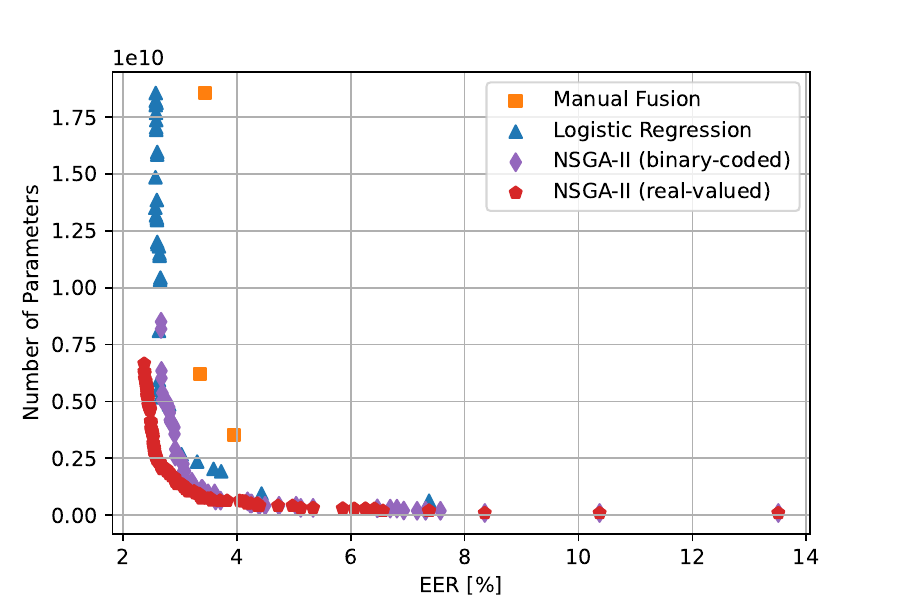}
    \caption{Multi-objective deepfake speech detector fusions comparing EER [\%] (horizontal axis) against number of parameters (vertical axis). The red and purple markers represent the global non-dominated solutions (super-Pareto front) found by 10 runs of the NSGA-II real-valued and binary variants, respectively. Single points for logistic regression (blue markers) and manual averaging (orange markers) are shown for comparison, representing baseline performance and complexity. Best viewed in color.}
    \vspace{-0.7em}
    \label{fig:results}
\end{figure*}

We compare the multi-objective evolutionary algorithm (NSGA-II) based fusion variants against traditional aggregation methods and logistic regression fusion. To ensure robust evaluation, each variant was executed for 10 independent runs. The final results in \autoref{fig:results} represent the super-Pareto front, which is constructed by aggregating the solutions from all independent runs and extracting the global non-dominated set.

The red markers illustrate the super-Pareto front obtained by the real-valued variant of NSGA-II, while the purple markers show the super-Pareto front of the binary variant. The real-valued variant generally demonstrates a superior trade-off, dominating the binary variant over most of the objective space. This indicates its ability to find fusion configurations that achieve lower EERs for a given number of parameters, or require fewer parameters for a comparable EER. Both fronts exhibit good coverage and spread, offering a wide range of compromise solutions from highly performant (lower EER, more parameters) to highly efficient (slightly higher EER, significantly fewer parameters). It is evident that the multi-objective optimization performed by NSGA-II provides a substantial advantage: both Pareto fronts generated by NSGA-II clearly dominate the solutions created by logistic regression or manual averaging.

Focusing on pure performance, the NSGA-II real-valued variant achieved the lowest EER of $2.37\%$ with approximately 6.6B parameters, representing an extreme point on its Pareto front where performance is heavily prioritized. The binary variant's best EER was slightly higher at $2.66\%$, further confirming the real-valued variant's slight edge in performance optimization. Of course, these results could be achieved with a simpler, single-objective evolutionary algorithm when focusing on pure performance at the expense of higher system complexity. For comparison, manual averaging yielded an EER of 3.35\% with approx.~6B parameters. The logistic regression fusion achieved an EER of 2.57\% with 18.56B parameters; however, a very similar EER (2.63\%) was achieved using logistic regression with only the top ten base detectors, i.e., with only 5.5B parameters.

\subsection{Comparison with State-of-the-Art}

To put our results into context, we compare them against the top-ranking teams of the ASVspoof 5 Challenge~\cite{wang24_asvspoof}. For completeness, we report both EER and minDCF~\cite{wang24_asvspoof} in the comparison. The systems represent the current state-of-the-art in deepfake speech detection. The winning submission \textit{T43}~\cite{chen24_asvspoof} employed a large-scale ensemble (approximately 5B parameters) and leveraged extensive optimization techniques for the underlying subsystems, including model fine-tuning, data augmentation, and additional external training data.

\autoref{tab:results} presents a comparison of our proposed multi-objective fusion variants against three state-of-the-art benchmarks. While the T43 winner achieved an EER of 2.59\% with approx.~5B parameters, the NSGA-II real-valued variant generated solutions that achieved a lower EER of 2.37\%. Critically, it also identified configurations with significantly fewer parameters, as evident from its Pareto front, even without fine-tuning and extensive optimization of the underlying detectors.

\begin{table}[htbp]
\caption{Comparison with ASVspoof 5 top teams. Selected real-valued NSGA-II solutions provide direct reference points to the teams in terms of either EER or parameter count.}
\small
\centering
\renewcommand{\arraystretch}{1.2}
\begin{tabular}{l c r r}
\hline
\textbf{System} & \textbf{EER} & \textbf{minDCF} & \textbf{Parameters} \\
\hline
\textit{T27} \cite{stourbe24_asvspoof} & 3.42\% & 0.0937 & approx. 286M \\
\textit{T23} \cite{aliyev24_asvspoof} & 3.41\% & 0.0936 & approx. 475M \\
\textit{T43} (Winner) \cite{chen24_asvspoof} & 2.59\% & 0.0750 & approx. 5B \\
\hline
Ours & 4.36\% & 0.1268 & 416M \\
Ours & 3.38\% & 0.0965 & 720M \\
Ours & 2.59\% & 0.0747 & \textbf{2.49B} \\
Ours & 2.43\% & 0.0706 & 4.86B \\
\textbf{Ours} & \textbf{2.37\%} & \textbf{0.0684} & 6.6B \\
\hline
\end{tabular}
\label{tab:results}
\vspace{-2em}
\end{table}

Importantly, the multi-objective nature of our approach enables efficient trade-offs between performance and complexity. Apart from the best-performing systems, we found a configuration that achieves an EER identical to that of the challenge winner (2.59\%) while utilizing only half of the parameters (2.49B). This demonstrates that evolutionary multi-objective fusion can achieve highly competitive performance while also offering substantially more resource-efficient system configurations.

\section{Discussion}

Our experiments highlight the efficiency and applicability of evolutionary optimization in constructing deepfake detection fusions. Importantly, we observed four interesting phenomena:

\begin{enumerate}
    \item Seeding the population with base detectors is important for faster convergence and conserving computational resources. This effectively injects domain knowledge into the optimization, allowing the algorithm to refine high-performing candidates immediately rather than spending generations to rediscover the baseline performance.

    \item Based on the discovered solutions, fusing a few smaller complementary detectors leads to systems with lower EER and fewer parameters compared to using a large monolithic system. 

    \item NSGA-II is able to discover configurations that yield comparable EER with fewer parameters when compared to other fusion strategies. This ability to optimize across multiple, often conflicting, objectives is a core strength of our approach, making it particularly well-suited for complex system design problems.

    \item The resulting set of Pareto-optimal solutions offers flexible deployment options. Users can select the optimal trade-off for their specific scenario.
\end{enumerate}

That being said, there are several limitations to our work. While effective, our approach currently relies on score-level fusion, which may overshadow deeper, feature-level interactions that joint fine-tuning could exploit. Additionally, the maximum performance is bounded by the properties of the frozen base detectors. The quality of the final fusion heavily depends on the underlying detectors. Optimizing the base detectors (where feasible) before fusion can greatly enhance the final performance. Future work can focus on investigating better-suited objectives (e.g., using inference latency directly instead of parameter count as a proxy), expanding our approach to multiple objectives, or examining end-to-end fine-tuning of the obtained fusion ensemble.

\section{Conclusion}

In this work, we introduce a multi-objective evolutionary framework for fusing deepfake speech detectors. By simultaneously optimizing for detection accuracy and system complexity, our approach eliminates the need for manual ensemble selection and effectively navigates the trade-off between raw performance and saving computational resources.

We demonstrated that the real-valued NSGA-II variant can leverage the diversity of frozen pre-trained models to construct ensembles that surpass the current state-of-the-art, achieving an EER of 2.37\% (compared to the ASVspoof 5 winner's 2.59\%). More importantly, the multi-objective approach enables the discovery of configurations that yield comparable performance with significantly smaller resource requirements.

Our framework provides a comprehensive set of Pareto-optimal solutions. This enables informed deployment decisions and the selection of a solution that fits both security requirements and specific hardware constraints.




\bibliography{IEEEabrv,references}
\bibliographystyle{IEEEtran}







\end{document}